\documentclass[thmsa,notitlepage,11pt]{article}
\usepackage{amsmath}
\usepackage{amssymb}
\usepackage{amsfonts}
\usepackage{sw20lart}
\usepackage[onehalfspacing]{setspace}

\input{tcilatex}

\begin{document}

\title{{\LARGE Vector Goldstone Boson and Lorentz Invariance\thanks{%
Talk presented at \textquotedblleft Miami 2004: A topical conference on
elementary particle physics and cosmology\textquotedblright\ Miami, Florida,
Dec 15-19, 2004.}}}
\author{{\large Ling-Fong Li} \\
{\small Department of Physics, Carnegie Mellon University, Pittsburgh, PA
15213}}
\maketitle

\begin{abstract}
Spontanous symmetry breaking usually gives spin 0 Goldstone bosons for the
case of internal symmetries and spin 1/2 fermions for the supersymmetry. The
spontaneous breaking of higher dimensional Lorentz symmetry can give vector
Goldstone boson in 4-dimension. 
\end{abstract}

\bigskip

\section{Introduction}

\bigskip

Spontaneous symmetry breaking has produced many interesting phenomena in
many different areas of physics. The celebrated Goldstone theorem (\cite%
{Goldstone}) implies the presence of zero energy excitation for the
spontaneous breaking of a continuous symmetry. In the context of
relativistic field theory, this gives rise to a massless particle, the
Goldstone particle. For the case of internal symmetries, the spontaneous
symmetry breaking gives massless spin zero mesons, the Goldtone bosons,
while the spontaneous breaking of supersymmetry yields massless spin 1/2
fermions, the Goldstinos. On the other hand, to have Goldstone boson with
spin 1 seems to require breaking of Lorentz symmetry. In this note, we
discuss the relation between the vector Goldstone boson and Lorentz
invariance. It turns out that in theories with extra dimensions (\cite{Extra}%
) it is possible to have vector Goldstone boson without breaking the
4-dimensional Lorentz symmetry (\cite{LFLI}) .

\bigskip

\section{Spin 0 and 1/2 Goldtone Particles}

We first illustrate the Goldstone theorem (\cite{Golstone}) to set the
framework for the discussion. Suppose the Lagrangian density is invariant
under a continuous transformation which gives rise to a conserved current,%
\begin{equation*}
\partial ^{\mu }J_{\mu }\left( x\right) =0.
\end{equation*}%
This implies\ a conserved charge which is the generator of the symmetry
transformation,%
\begin{equation*}
Q=\int J_{0}\left( x\right) \,d^{3}x,\qquad \dfrac{dQ}{dt}=0
\end{equation*}%
Suppose that there are two local operators $A$ and $B,$ which are related by
the generator of the symmetry transformation $Q$,%
\begin{equation}
\left[ Q,\;A\right] =B.  \label{commutator}
\end{equation}%
If for some dynamical reasons, the operator $B$ has non-zero vacuum
expectation value (VEV),%
\begin{equation}
\left\langle 0\left\vert B\right\vert 0\right\rangle \neq 0,  \label{SSB}
\end{equation}%
we say that the symmetry is broken spontaneously. It follows from Eq(\ref%
{SSB}) that 
\begin{equation}
Q|0>\neq 0
\end{equation}%
i. e. the vacuum is no longer invariant under the symmetry transformation.
The Goldstone theorem then states that there exists a state $|g\left( 
\overset{\rightarrow }{p}\right) >$ with the property that its energy 
\begin{equation*}
E_{g}\left( \overset{\rightarrow }{p}\right) \rightarrow 0,\qquad \text{as \
\ \ \ \ }\overset{\rightarrow }{p}\rightarrow 0.
\end{equation*}%
Here $\overset{\rightarrow }{p}$ is the momentum of the particle. From the
relativistic energy momentum relation, $E=\sqrt{\overset{\rightarrow }{p}%
^{2}+m^{2}}$ this implies the existence of massless particle. Note that the
Goldstone state $|g\left( \overset{\rightarrow }{p}\right) >$ couples
directly to the operators, $A$ and $J_{\mu },$%
\begin{equation*}
\left\langle 0\left\vert J_{\mu }\right\vert g\left( \overset{\rightarrow }{p%
}\right) \right\rangle \neq 0,\qquad \left\langle 0\left\vert A\right\vert
g\left( \overset{\rightarrow }{p}\right) \right\rangle \neq 0.
\end{equation*}%
This means that the Goldstone state $|g\left( \overset{\rightarrow }{p}%
\right) >$ is also the quantum of the local operator $A.$ Note that the
symmetry breaking condition in Eq (\ref{SSB}) implies that whatever the
quantum number carried by the operator $B$ will not be conserved by the
vacuum.

\qquad A simple example of spin 0 Goldstone boson is provided by the
spontaneous breaking of the chiral symmetry in the low energy hadron physics
(\cite{Chiral}). Here chiral charges $Q_{5}^{a}$ transform pion fields $\pi
^{b}$ into scalar $\sigma $ field,%
\begin{equation}
\left[ Q_{5}^{a},\;\pi ^{b}\right] =\delta _{ab}\sigma .
\label{chiral transform}
\end{equation}%
Suppose the effective interaction between $\pi $ and $\sigma $ is of the
form,%
\begin{equation}
V_{eff}=-\dfrac{\mu ^{2}}{2}\left( \overset{\rightarrow }{\pi }^{2}+\sigma
^{2}\right) +\dfrac{\lambda }{4}\left( \overset{\rightarrow }{\pi }%
^{2}+\sigma ^{2}\right) ^{2}  \label{effv}
\end{equation}%
This is invariant under chiral $SU\left( 2\right) _{L}\times SU\left(
2\right) _{R}$ transformations. The minimization of $V$ $_{eff}$ leads to
non-zero VEV\ for $\sigma ,$%
\begin{equation}
\left\langle 0\left\vert \sigma \right\vert 0\right\rangle =\sqrt{\dfrac{\mu
^{2}}{\lambda }}\neq 0  \label{condensate}
\end{equation}%
In this case, the pion fields $\overset{\rightarrow }{\pi }$ , which are
partners of $\sigma $ under the chiral transformation, Eq (\ref{chiral
transform}), are the massless spin zero Goldstone bosons. From the point of
view of more fundamental theory QCD, $\sigma $ field is just the quark
bilinear $\overset{\_}{q}q$ and Eq (\ref{condensate}) is equivalent to%
\begin{equation*}
\left\langle 0\left\vert \overset{\_}{q}q\right\vert 0\right\rangle \neq 0
\end{equation*}%
and the Goldstone bosons are bound states,%
\begin{equation*}
\pi ^{a}\symbol{126}\overset{\_}{q}\gamma _{5}\tau ^{a}q
\end{equation*}

For the case of spontaneous breaking of supersymmetry, there are two
possibilities.\ One is generated by the chiral superfield  where we have the
anti-commutation relation,%
\begin{equation*}
\left\{ Q_{\xi },\,\psi \right\} =\sqrt{2}F.
\end{equation*}%
Here $Q_{\xi }$ is the supercharge which transform fermions into bosons and
vice versa, $\psi $ is the fermion component and $F$ the scalar auxiliary
component of the chiral superfield. If $F$ develops non-zero VEV (\cite%
{F-term}),%
\begin{equation*}
\left\langle 0\left\vert F\right\vert 0\right\rangle \neq 0
\end{equation*}%
then fermion $\psi ,$ becomes a Goldstone particle ( Goldstino) and $\psi $
is also a partner of a massive physical scalar field. In the case of
spontaneous breaking by a vector superfield, the anti-commutatuion relation
of interest is, 
\begin{equation*}
\left\{ Q_{\xi },\,\lambda \right\} =\sqrt{2}D
\end{equation*}%
where $\lambda $ is the fermion component and $D$ is the auxiliary scalar
component of the vector superfield. When $D$ develops VEV (\cite{D-term}), 
\begin{equation*}
\left\langle 0\left\vert D\right\vert 0\right\rangle \neq 0
\end{equation*}%
then $\lambda $ is the Goldstone fermion and is a partner of a massive
physical vector particle.

\section{Vector Goldstone Boson}

In all the cases discussed above, the fields which develops non-zero VEV\ \
all have spin 0 and the Goldstone particles are partners of these fields
under the symmetry transformations. In the usual 4-dimensional field theory,
the vector field is not related to any scalar field. The only way to get
vector Goldstone boson is to break the Lorentz invariance spontaneoulsy, For
example,\ in analogy with Eq (\ref{commutator}), suppose there are two
vector fields which are related by some symmetry transformation,%
\begin{equation*}
\left[ Q,\,A^{\mu }\right] =B^{\mu }
\end{equation*}%
If $B^{\mu }$ develops VEV,%
\begin{equation}
\left\langle 0\left\vert B^{\mu }\right\vert 0\right\rangle \neq 0
\label{ssb-v}
\end{equation}%
then the quantum of $A^{\mu }$ field, will be a spin 1 massless Goldstone
particle. Clearly, the symmetry breaking condition in Eq (\ref{ssb-v})
breaks the Lorentz invariance.. Even though there are stringent experimental
limits on the possible violation of Lorentz symmetry, recently there have
been renew interest in studying this issue. However, it has been pointed out
(\cite{LFLI}) that in theories with extra dimensions, it is possible to have
vector Goldstone boson without breaking the Lorentz symmetry in 4-dimension.
As an example (\cite{LFLI}), consider a vector field in 5-dimensional
theory, $\phi _{A},$ $A=0,1,2,3,4.$ In analogy to Eq (\ref{effv}), we can
write down an effective interaction of the form,%
\begin{equation*}
V\left( \phi \right) =\dfrac{\mu ^{2}}{2}\left( \phi _{A}\phi ^{A}\right) +%
\dfrac{\lambda }{4}\left( \phi _{A}\phi ^{A}\right) ^{2}.
\end{equation*}%
For the case, $\mu ^{2}>0,$ we can choose 
\begin{equation}
\left\langle 0\left\vert \phi _{4}\right\vert 0\right\rangle =\sqrt{\dfrac{%
\mu ^{2}}{\lambda }}  \label{symmbre condi}
\end{equation}%
to minimize the potential $V\left( \phi \right) .$ This will break the
Lorentz symmetry in 5-dimension, $SO(4,1)$ to that of 4-dimension, $SO(3,1).$
As a consequence, the 4-dimensional vector fields $\phi _{\mu },$ $\mu
=0,1,2,3$ which are partners of $\phi _{4},$ are massless, the vector
Goldstone boson. Note that symmetry breaking condition in Eq (\ref{symmbre
condi}) can come from fermion condensate,%
\begin{equation*}
\left\langle 0\left\vert \overset{\_}{\psi }\gamma _{A}\psi \right\vert
0\right\rangle =v\delta _{A4}
\end{equation*}%
In this case the vector Goldstone boson correspond to the composite fields $%
\overset{\_}{\psi }\gamma _{\mu }\psi ,$ $\mu =0,1,2,3.$

\section{Goldstone Photon}

It was originally explored by Bjorken forty years ago (\cite{Bjorken}) the
possibility  that the photon is massless because of spontaneous symmetry
breaking rather gauge invariance. This idea has been revisited more recently
by Bjorken and several other authors (\cite{GPhoton}). Even though this idea
does not seem to be very attractive in view of the spectacular success of
gauge theories in recent years, it is of interest to study whether this idea
is phenomenological viable. As we have mentioned before, in 4-dimension to
have photon as vector Goldstone boson requires the breaking of Lorentz
invariance. We will now discuss a simple version of this scheme to
illustrate the idea. The starting point is to write down a self-interacting
fermion field theory of the form,%
\begin{equation}
L=\overset{\_}{\psi }\left( i\gamma ^{\mu }\partial _{\mu }-m\right) \psi
+\sum_{n=1}^{\infty }\lambda _{2n}\left( \overset{\_}{\psi }\gamma _{\mu
}\psi \right) ^{2n}  \label{Nambu-J}
\end{equation}%
where $\lambda _{2n}$ 's are coupling constants. This resembles the
Nambu-Jona-Lasino type of theory (\cite{Nambu}) for the spontaneous symmetry
breaking. One can introduce an auxiliary field $A^{\mu }$ to rewrite the
Lagrangian as,%
\begin{equation*}
L=\overset{\_}{\psi }\left( i\gamma ^{\mu }\partial _{\mu }-m\right) \psi
+eA^{\mu }\overset{\_}{\psi }\gamma _{\mu }\psi -V\left( A^{\mu }A_{\mu
}\right) 
\end{equation*}%
where the effective potential $V$ is 
\begin{equation*}
V\left( A^{2}\right) =\Lambda ^{4}\sum_{n=1}^{\infty }V_{n}\left( \dfrac{%
A^{2}}{\Lambda ^{2}}\right) ^{n}\qquad \text{with \ \ }A^{2}=A^{\mu }A_{\mu }
\end{equation*}%
Here $V_{n}$'s are dimensionless constants related to $\lambda _{2n}$ in Eq (%
\ref{Nambu-J}) and $\Lambda $ is some parameter with dimension of mass. This
potential can generate a non-zero VEV of the form,%
\begin{equation*}
\left\langle 0\left\vert A_{\mu }\right\vert 0\right\rangle =c\Lambda n_{\mu
},\qquad 
\end{equation*}%
Here $c$ is some dimensionless constant and $n_{\mu }$ a space-like unit
vector. This breaks the Lorentz symmetry and give 3 massless Goldstone modes
in which two of them are the transverse photons and the other one is a
time-like mode. If $\Lambda $ is very large, at energies small compared with 
$\Lambda $ we can have approximate Lorentz symmetry and any deviation can be
suppressed by making $\Lambda $ very large enough.

The possible violation of Lorentz invariance has also been investigated in
the studies of string theory and gravitational interaction (\cite{String}).
The possibility that the graviton is also a Goldstone boson has been
discussed in the literature(\cite{G-graviton}).


\begin{thebibliography}{99}
\bibitem{Goldstone} J. Goldstone, \emph{Nuovo Cimento }\textbf{19, }154
(1961); J. Goldstone, A. Salam, and S. Weinberg, \emph{Phys. Rev.} \textbf{%
127, }965 (1962).

\bibitem{Nambu} Y. Nambu, \emph{Phys Rev. Lett}. \textbf{4}, 380 (1960), Y.
Nambu and G. Jona-Lasinio. \emph{Phys. Rev}. \textbf{122, }345, (1961).

\bibitem{SUSY} See for example, J Wess and J. Bagger, " \emph{Supersymmetry
and Supergravity} " Second edition, Princeton University Press, (1992); S.
Weinberg, " \emph{Quantum Theory of Fields}", Vol 3, Cambridge University
Press, (2000).

\bibitem{Chiral} M. Gell-Mann and M. Levy, \emph{Nuovo Cimento}, \textbf{16, 
}705 (1960).

\bibitem{LFLI} Ling-Fong Li, \emph{Eur.Phys.J. }\textbf{C28}:145-146, (2003)

\bibitem{Extra} N. Arkani-Hamed, S. Dimopoulos, and G. Dvali, \emph{Phys.
Lett}. \textbf{B429, }263 (1998); L. Randall and R. Sundrum, \emph{Phys.
Rev. Lett}. \textbf{83, }3370 (1999).

\bibitem{D-term} P. Fayet and J. Iliopoulos, \emph{Phys. Lett}. \textbf{51B}%
, 461 (1974).

\bibitem{F-term} L. O'Raieartaigh, \emph{Nucl Phys}., \textbf{B96}, 331
(1975).

\bibitem{Bjorken} J. D. Bjorken, \emph{Ann. Phys.} (N. Y. ) \textbf{24}, 174
(1963).

\bibitem{GPhoton} J. D. Bjorken, hep-th/0111196, P Kraus and E. T.
Tomboulis, \emph{Phys. Rev.} \textbf{D66}, 045015(2002), \ Alejandro
Jenkins, \emph{Phys. Rev. }\textbf{D69} : 105007 ( 2004). J. L. Chkareuli,
C.D. Froggatt, R. N. Mohapatra, and H. B. Nielsen, hep-th/0412225, "Photon
as Goldstone Boson: Nonlinear $\sigma $ Model for QED".

\bibitem{G-graviton} M. J. Duff, \emph{Phys Rev }\textbf{D12}  3969 (1975),
P Kraus and E. T. Tomboulis, \emph{Phys. Rev}. \textbf{D66}, 045015 (2002).

\bibitem{String} N. Arkani-Hamed, H. C. Cheng, M. A. Luty and S. Mukoyama, 
\emph{JHEP} \textbf{0405}, 074 (2004); \ B. M. Gripaios, \emph{JHEP} \textbf{%
0410}, 069 (2004); T. Jacobson, S. Liberati, and D. Mattingly arXiv
:gr-qc/0404067.

\bibitem{Lorentz}  V.A.\ Kostelecky, \ and S. Samuel \emph{Phys. Rev}.%
\textbf{\ D39}, 683 (1989).
\end{thebibliography}
\end{document}